\documentclass[journal]{IEEEtran}

\usepackage{amsmath, amssymb}
\usepackage{multirow}
\usepackage{flushend}
\usepackage{epstopdf}
\usepackage{graphics}
\usepackage{tabularx}
\usepackage{array}
\usepackage{colortbl}
\usepackage{rotating}
\usepackage{placeins}
\usepackage[table]{xcolor}
\usepackage{booktabs}
\usepackage{graphicx}
\usepackage{subfig}
\usepackage{array}
\usepackage{algorithm}
\usepackage{algorithmic}
\usepackage{enumitem}
\usepackage{times}
\usepackage{epstopdf}
\usepackage{longtable}
\usepackage{booktabs}
\usepackage{float}
\usepackage{tabulary}
\usepackage{multirow}
\usepackage{graphicx}
\usepackage{epstopdf}
\usepackage{comment}
\usepackage{enumitem}
\usepackage{tabulary}
\usepackage{balance}

\def\0{{\bf 0}}

\makeatletter
\renewcommand{\paragraph}{%
	\@startsection{paragraph}{4}%
	{\z@}{1.00ex \@plus 1ex \@minus .2ex}{-1em}%
	{\normalfont\normalsize\bfseries}%
}
\makeatother
\newcolumntype{L}[1]{>{\raggedright\let\newline\\\arraybackslash\hspace{0pt}}m{#1}}
\newcolumntype{C}[1]{>{\centering\let\newline\\\arraybackslash\hspace{0pt}}m{#1}}
\newcolumntype{R}[1]{>{\raggedleft\let\newline\\\arraybackslash\hspace{0pt}}m{#1}}
\newcolumntype{H}{>{\setbox0=\hbox\bgroup}c<{\egroup}@{}}

\let\OLDthebibliography\thebibliography
\renewcommand\thebibliography[1]{
  \OLDthebibliography{#1}
  \setlength{\parskip}{0pt}
  \setlength{\itemsep}{0pt plus 0.3ex}
}

\begin{document}
\title{Personality, Culture, and System Factors - Impact on Affective Response to Multimedia}

\author{Sharath Chandra Guntuku, Michael James Scott,  \\
	Gheorghita Ghinea ~\IEEEmembership{Member,~IEEE,} Weisi Lin~\IEEEmembership{Fellow,~IEEE}
	\thanks{Sharath Chandra Guntuku and Weisi Lin are with School of Computer Engineering, Nanyang Technological University, Singapore, 639798 E-mail: sharathc001@e.ntu.edu.sg, wslin@ntu.edu.sg .}
	\thanks{Michael James Scott is with the Games Academy, Falmouth University, Cornwall, United Kingdom E-mail: michael.scott@falmouth.ac.uk.}
	\thanks{Gheorghita Ghinea is with Department of Computer Science, Brunel University, London, United Kingdom E-mail: george.ghinea@brunel.ac.uk.}
	\thanks{Manuscript received .}}

\markboth{Preprint}%
{Authors \MakeLowercase{\textit{et al.}}: }

\maketitle

\begin{abstract}
	Whilst affective responses to various forms and genres of multimedia content have been well researched, precious few studies have investigated the combined impact that multimedia system parameters and human factors have on affect. Consequently, in this paper we explore the role that two primordial dimensions of human factors - personality and culture - in conjunction with system factors - frame rate, resolution, and bit rate - have on user affect and enjoyment of multimedia presentations. To this end, a two-site, cross-cultural study was undertaken, the results of which produced three predictve models. Personality and Culture traits were shown statistically to represent 5.6\% of the variance in positive affect, 13.6\% in negative affect and 9.3\% in enjoyment. The correlation between affect and enjoyment,  was significant. Predictive modeling incorporating human factors showed about 8\%, 7\% and 9\% improvement in predicting positive affect, negative affect and enjoyment respectively when compared to models trained only on system factors. Results and analysis indicate the significant role played by human factors in influencing affect that users experience while watching multimedia.
\end{abstract}

\section{Introduction}
\IEEEPARstart{M}{u}ltimedia content produces diverse affective (emotional) responses in humans. When warmth and competence shape our judgements of people and organizations, and when perceived together they cause active behavioral responses from the viewers \cite{fiske2007universal}. Daily we witness several organizations put forward their missions in the form of ad campaigns. While most of these ads fail to attract our attention, some of them leave a lasting impression in our minds. Take the example of the campaign by Volvo, which was listed as one of most unforgettable ad campaigns of 2013 \cite{forbesad} or Singapore's Ministry of Education `Teach' campaign. The huge success of such ad campaigns is attributed to how story-telling components are shaped into emotion-evoking communication, structured to stimulate action. 

Ad campaigns are but one specific scenario which illustrate the importance and challenge of modeling multimedia-evoked emotions. Publicity campaigns, movies, sports, educational material, games, to name a few, all require research into investigating a user's Quality of Experience (QoE) \cite{jain2004quality,  wu2009quality}, of which affect is an important dimension. Experience of affect is defined as the positivity of emotions that viewers feel while watching videos. The problem is not just limited to content- or genre-based analysis of multimedia. This is because a video which arouses a positive emotion in one person might arouse a negative emotion in the other (depending on the nature of content and users' cultural and psychophysical frameworks which influence their perception) \cite{winter1997individual}. Whilst this is understood, how system parameters impact on the affective experience of those viewing multimedia content remains largely unexplored. What is also relatively unexplored is whether, and if so, to what degree, human factors also impact upon affective responses. These are the two main issues which we address in this paper - does multimedia content and system quality parameters with which it is presented evoke different affective responses depending on an individual's personality and culture? 

Answering this question involves understanding the subjective nature of emotions and how crucial a role human factors play in modeling experience of affect (emotion), thereby addressing users' needs for emotion-sensitive video retrieval \cite{choe2013estimating}. In this work, we attempt to understand how personality \cite{matthews2003personality} and culture \cite{hoftede2010cultures} influence users' experience of affect and enjoyment in multimedia. Specifically, the following research questions are posed:
\setlist[enumerate,1]{leftmargin=1.5cm}
\begin{enumerate}[label=RQ {\arabic*}.]
	\item Can a model based on multimedia system characteristics (Bit-Rate, Frame-Rate and framesize) and human factors (i.e., personality and culture) predict the intensity of affect (both positive and negative) and enjoyment?
	\item Which system characteristics and human factors influence the experience of affect and enjoyment the most?	
	\item What is the relationship between experience of affect (both positive and negative) and enjoyment across stimuli? 
	\item How do predictive models perform on the task of automatic assessment of experience of affect and enjoyment of videos?
\end{enumerate}

By investigating how different dimensions of these human factors modulate users' experience of affect and enjoyment, and specifically by understanding the correlation between enjoyment and perception of affect, we intend to provide initial findings for multimedia content creators to achieve maximal user satisfaction with respect to the content they create and deliver to diverse users. 

\begin{table*}[t!]
	\centering
	\caption{Datasets for Affective Modeling of Videos: Most of them implicitly assume that, given a video, the affect experienced by different users will be the same.}
	\label{tab:datasets}
	\renewcommand{\arraystretch}{1.2}
	\begin{tabular}{C{5.3cm}|C{3cm}C{.75cm}C{1cm}C{2.8cm}C{1cm}HC{1cm}}
		\hline
		\textbf{Dataset}                                                                                               & \multicolumn{1}{c|}{\textbf{Category}} & \multicolumn{1}{C{.75cm}|}{\textbf{\# Videos}} & \multicolumn{1}{C{1cm}|}{\textbf{\# Annotators}} & \multicolumn{1}{C{3.3cm}|}{\textbf{Type of Annotation}}                                                                     & \multicolumn{1}{C{1cm}|}{\textbf{Users' Profile}}        & \multicolumn{1}{H}{\textbf{Multiple emotions on same video}} & \textbf{Enjoyment} \\ \hline
		Mutual Information-Based Emotion Recognition  \cite{cui2013mutual}                                                                 & 2 Mainstream Movies                          & 655                                        & -                                              & One value each (0,1) for valence and arousal                                                                         & No                                                           & No                                                            & No                       \\ \cline{1-1} \hline
		Predicting Emotions in User-Generated Videos \cite{jiang2014predicting} & User Generated Videos                          & 1101                                        & -                                              & Plutchik's emotions used as search keywords on Youtube and Flickr                                                                        & No                                                           & No                                                            & No                       \\ \cline{1-1} \hline
		A Connotative Space for Supporting Movie Affective Recommendation \cite{benini2011connotative}                                              & 25 Mainstream Movies                         & 25                                         & 240                                            & Warmth of the scene atmosphere, Dynamic pace of the scene, Energetic impact & No                                                           & No                                                            & No                       \\ \cline{1-1} \hline
		Music Video Affective Understanding Using Feature Importance Analysis  \cite{cui2010music}                                        & Multilingual Music Videos                    & 250                                        & 11                                             & 4 point valence, arousal                                                                                             & No                                                           & No                                                            & No                       \\ \cline{1-1} \hline
		Utilizing Affective Analysis for Efficient Movie Browsing  \cite{zhang2009utilizing}                                                   & 13 Mainstream Movies                         & 4000                                       & -                                              & One value each (0,1) for valence and arousal                                                                         & No                                                           & No                                                            & No                       \\ \cline{1-1} \hline
		Affective visualization and retrieval for music video      \cite{zhang2010affective}                                                    & Music Videos                                 & 552                                        & 27                                             & One value each (0,1) for valence and arousal                                                                         & Individual user's ratings& No                                                            & No                       \\ \cline{1-1} \hline
		Affective level video segmentation by utilizing the pleasure-arousal-dominance information  \cite{arifin2008affective}                   & 13 Maintream Movies                          & 43                                         & 14                                             & Ekman's 6 Emotions (1-10)                                                                                            & No                                                           & Yes                                                           & No                       \\ \cline{1-1} \hline
		Emotional identity of movies \cite{canini2009emotional}                                                                                  & 87 Mainstream Movies                         & 87                                         & -                                              & First 2 Genres from IMDB                                                                                             & No                                                           & -                                                             & No                       \\ \cline{1-1} \hline
		Affective audio-visual words and latent topic driving model for realizing movie affective scene classification \cite{irie2010affective} & 24 Mainstream Movies                         & 206                                        & 16                                             & Plutchik's 8 emotions (1-7)                                                                                          & No                                                           & No                                                            & No                       \\ \cline{1-1} \hline
		Determination of emotional content of video clips by low-level audiovisual features \cite{teixeira2012determination}                           & 24 Mainstream Movies                         & 346                                        & 16                                             & Pleasure Arousal Dominance values (1-7)                                                                              & No                                                           & No                                                            & No                       \\ \cline{1-1} \hline
		LIRIS-ACCEDE   \cite{baveye2013large}                                                                                                & 160 Creative Commons Movies                  & 9800                                       & 1517                                           & Rating-by-Comparison on Valence-Arousal                                                                              & No                                                           & No                                                            & No                       \\ \cline{1-1} \hline
		FilmStim  \cite{schaefer2010assessing}                                                                                                     & 70 Mainstream Movies                         & 70                                         & 364                                            & 24 emotional classification criteria                                                                                 & No                                                           & No                                                            & No                       \\ \cline{1-1} \hline
		MediaEval \cite{reuter2013social}                                                                                                     & Travelogue series & 126                                        & -                                              & Popularity/Boredom                                                                                                   & No                                                           & No                                                            & No                       \\ \cline{1-1} \hline
		Content-based prediction of movie style, aesthetics andaffect: Data set and baseline experiment \cite{tarvainen2014content}               & 14 Mainstream Movies                         & 14                                         & 73                                             & ValenceArousal                                                                                                       & No                                                           & No                                                            & No                       \\ \cline{1-1} \hline
		DEAP    \cite{koelstra2012deap}                                                                                                       & Music Videos                                 & 40                                         & 32                                             & PhysiologicalSignals, FaceVideos, Valence, Arousal                                                                       & No                                                           & No                                                            & \textbf{Yes}             \\ \cline{1-1} \hline
		MAHNOB HCI   \cite{soleymani2012multimodal}                                                                                              & Mainstream Movies                            & 20                                         & 27                                             & ValenceArousal, 6 emotion categories                                                                                 & No                                                           & No                                                            & No                       \\ \cline{1-1} \hline
		\textbf{CP-QAE-I (Our Dataset)}                                                                                              & \textbf{14 Mainstream Movies}                & \textbf{144}                               & \textbf{114}                                   & \textbf{16 sets of emotion-related adjectives from DES}                                                              & \textbf{Yes}                                                 & \textbf{Yes}                                                  & \textbf{Yes}             \\ \hline
	\end{tabular}
\end{table*}

\section{Related Work}

There are several studies which aim to predict affective responses to multimedia (see \cite{calvo2010affect, wangvideo, zeng2009survey} for a thorough review). Some focus on distilling the influence of specific cinematographic theories \cite{canini2013affective}, types of segment and shot \cite{bhattacharya2013towards}, the use of colour \cite{wei2004color} and connotative space \cite{benini2011connotative}. Apart from the works mentioned above, there has been research focused on modeling the different audio-visual features to predict emotions \cite{li2015horror, metallinou2012context, schuller2011avec, soleymani2009bayesian, sun2007video, teixeira2012determination}. The features used in this work are inspired by those used in the literature, along with certain content-based descriptors which have been shown to perform well in several content understanding tasks \cite{borth2013sentibank,zhou2014learning}.  

Research on modeling emotional response in videos also often takes into account the facial expressions of viewers \cite{chiranjeevi2015neutral, navarathna2014predicting, shojaeilangari2015robust, wang2015micro, siddiqi2015human} and a range of complementary sensors (e.g., heart rate, EEG) to help measure the evoked emotions \cite{han2013representing,  koelstra2012deap, soleymani2012multimodal}. However, the extent to which physiological responses capture the subjective intensity of affect (which varies as a consequence of users' innate psychology) is unclear.

A consequence of this is that such studies implicitly assume that, given a video, the affect experienced by different users will be more or less the same. This is equally the case with affective video datasets (as seen in Table \ref{tab:datasets}). However prior research shows that individual differences can lead to varied experiences \cite{winter1997individual}. To illustrate this, evidence reveals a complex relationship between affective video tagging and physiological signals \cite{abadi2013user, choe2013estimating, wache2014secret}. As such, it is important to consider the subjective nature of affective perception. However, it is to be noted that we do not aim at creating a large-scale video dataset for affective modeling, rather our aim is to understand the influence of users' individual traits on their perception of affect and consequently develop a dataset towards this goal.

Personality can be a good tool to explore the systematic differences in users' individual traits \cite{matthews2003personality}. One popular model is the Five Factor Model (FFM) \cite{goldberg1990alternative}.

Certain traits are considerably influenced by the cultural background to which an individual belongs. Shared conceptions and collective norms characterize a local environment, and thereby shape the perception and cognition of those who associate with it. Differences in culture have been studied by Hofstede et al \cite{hoftede2010cultures}. Six cultural traits constitute the model --  masculinity, individualism, uncertainty avoidance index,  pragmatism, power distance, and indulgence.

Both human factors targeted in our study, namely personality and culture, are shown to reliably capture individual differences in multiple domains like language \cite{argamon2005lexical}, intonation of voice while speaking \cite{mohammadi2011humans, mohammadi2012automatic}, kind of photos one likes \cite{guntuku2015personality}, type of people one befriends \cite{golbeck2011predicting}, etc. (see \cite{vinciarelli2014survey} for a thorough review). Other examples include preference of genre for language learning in different cultures \cite{barza2014movie} and the respective cultural acceptance of some movie content \cite{craig2005culture} etc. Due to the consistency shown between these human factors and user behaviors, we use them to study how they influence users' experience of affect and enjoyment in multimedia.

\section{Data Collection}

To address the concern of modeling the influence of individual differences on the experience of affect and enjoyment, we build a dataset using videos which are annotated by users with diverse personality and cultural traits. This section describes the videos, the procedure to used to collect annotations and the descriptive statistics. 

\subsection{Video Dataset}
\label{subsec:dataset}
This study uses the CP-QAE-I dataset \cite{guntukucp} (\textbf{{http://1drv.ms/1M1bnwU}}), which consists of 12 purposively selected short clips from popular movies to cover different affective categories \cite{schaefer2010assessing}. Clips from a wide range of valence but low variance on arousal \cite{schaefer2010assessing} have been adopted to reduce  content-based biases, and the clips, their description and the means of  positive and negative affect are given in Table \ref{tab:content_resp} for the reader's reference. The content parameter also varies in cinematographic schemes utilized in the original movie production. There are three video coding parameters, namely bitrate -- with the settings 384kbps and 768kbps, resolution (frame-size) -- with the settings 480p and 720p, and framerate -- with the settings 5fps, 15fps and 25fps, and these result in twelve coding-quality combinations. As a result, the dataset contains 144 (12*12) video sequences.  

From the analysis with G*Power 3 \cite{faul2007g}  using the $F$-statistic as well as repeated measures, the  minimum required sample size is 64, utilizing the conventional error probabilities ($\alpha = 0.05, \beta = 0.2$) and with the assumption of existing medium effects ($f = 0.39$) with $r = 0.8$ correlation. 

\begin{table*}[t!]
	\begin{center}
		\caption{Marginal means of perceived responses (affect and enjoyment) on clips, after fixing the co-variates} 
		\label{tab:content_resp}
		\renewcommand{\arraystretch}{1.2}
		\setlength\tabcolsep{3pt}
		\begin{tabulary}{1\columnwidth}{lL{8cm}rrrr}
			\hline \noalign{\smallskip}
			MovieClip (Duration in Mins:Secs)     & Description from \cite{schaefer2010assessing}      & +ve Affect & -ve Affect & Enjoyment \\
			\noalign{\smallskip}
			\hline
			\noalign{\smallskip}
			A\_FISH\_CALLED\_WANDA (2:56) & 
			\textit{One of the characters is found naked by the owners of the house} & 0.184           & -0.536          & -0.037    \\
			AMERICAN\_HISTORY\_X (1:06) & \textit{A neo-Nazi kills an African-American man, smashing his head on the curb}  & -0.397          & 0.756           & -0.607    \\
			CHILDS\_PLAY\_II (1:07)     & \textit{Chucky beats Andy’s teacher
				with a ruler}  & -0.231          & 0.698           & -0.158    \\
			COPYCAT (1:04)        & \textit{One of the characters gets caught by a murderer in a toilet}        & -0.33           & 0.418           & -0.315    \\
			DEAD\_POETS\_SOCIETY\_1 (2:34) & \textit{A schoolboy commits suicide} & -0.331          & 0.341           & -0.504    \\
			DEAD\_POETS\_SOCIETY\_2 (2:23) & \textit{All the students climb on their desks to express their solidarity with
				Mr Keating, who has just been fired} & 1.053           & -0.553          & 0.725     \\
			FOREST\_GUMP (1:47)        & \textit{Father and son are reunited}   & 0.992           & -0.523          & 0.656     \\
			SE7EN\_1 (1:39)          & \textit{By the end of the movie, Kevin Spacey tells Brad Pitt that he beheaded
				his pregnant wife}     & -0.346          & 0.248           & 0.42      \\
			SE7EN\_3 (0:24)    & \textit{Policemen find the body of a man tied to a table}           & -0.431          & 0.03            & -0.306    \\
			SOMETHING\_ABOUT\_MARY (2:00)  & \textit{Mary takes sperm from Ted’s hair mistaking it for hair gel} & 0.468           & -0.72           & 0.471     \\
			THE\_PROFESSIONAL (2:44)   & \textit{The two main characters are separated forever}   & -0.194          & 0.216           & 0.254     \\
			TRAINSPOTTING  (0:40)      & \textit{The main character dives into a filthy toilet}   & -0.477          & -0.389          & -0.654  \\ 
			\hline
		\end{tabulary}
	\end{center}
	\small
	\raggedright{\textit{Based on estimated marginal means of a mixed-effects regression model. Covariates in the model are evaluated at the following values:  AGREEABLENESS = 7.45; EXTRAVERSION = 5.42; CONSCIENTIOUSNESS = 6.59; OPENNESS = 6.77; NEUROTICISM = 5.67; POWER DISTANCE = -34.29; MASCULINITY = -6.73; INDIVIDUALISM = 22.44; UNCERTAINTY AVOIDANCE = 40.83; INDULGENCE = -11.60 PRAGMATISM = 22.82;.}}
\end{table*}

\begin{table}[t!]
	\centering
	\small
	\caption{ BFI-10 questionnaire and associated Personality Traits: Each question is associated to a Likert scale. Questions are taken from \cite{goldberg1990alternative}.}
	\label{tab:bfi10}
	\begin{tabular}{|c|c|} \hline
		\textbf{Question} & \textbf{Trait} \\
		\hline
		I have few artistic interests & O\\
		I have an active imagination & O\\
		I tend to find fault with others & A\\
		I am generally trusting & A\\
		I get nervous easily & N\\
		I am relaxed, handle stress well & N\\
		I am reserved & E \\
		I am outgoing, sociable & E\\
		I tend to be lazy & C\\
		I do a thorough job & C\\
		\hline
	\end{tabular}
	\vspace{-0.2cm}
\end{table}

\subsection{Procedure}
The participants for data collection were 57 college students from each of the two universities with which the authors are affiliated (so totally 114 participants): 43 from Britain, 22 from India, 16 from China, 15 from Singapore and 18 from other nationalities. 28.9\% of the participants were female and 23.9 years  was the mean age, $\sigma = 3.68$. The corresponding cultural and personality traits are given  in Table \ref{tab:descstat}.

We applied a lab-based subjective testing approach. A set of videos were located locally on servers at the authors' universities. Users answered an online questionnaire from the local server (to avoid any latency issues over the Internet). Each user saw all 12 clips (in a random order), with different system characteristics and rated the experience of affect and their enjoyment of each sequence by completing questions immediately after viewing each. Informed consent and anonymity were assured at every stage of the study.  

Since human factors are studied, we aim to maximize ecological validity in recording users' viewing behavior, so there was no limit on time to finish the survey. However, owing to the nature of such studies, some participants dropped out.

Participants started the survey by answering the BFI-10 \cite{gosling2003very} and the VSM-2013 \cite{hoftede2010cultures} for assessment of personality and cultural traits. Afterwards, they were shown 12 videos under test, and were expected to give their ratings on all sequences. 73.7\% of the 114 participants did so. However, all participants rated a minimum of 3 videos, with the average being 10.8 ($\sigma = 2.56$). Over all, 1,232 ratings were collected ($90\%$ of the maximum possible).

\subsection{Measures}
\subsubsection{Positive and Negative Affect}  
\begin{figure}[tp!]
	\centering
	\includegraphics[width=1\columnwidth]{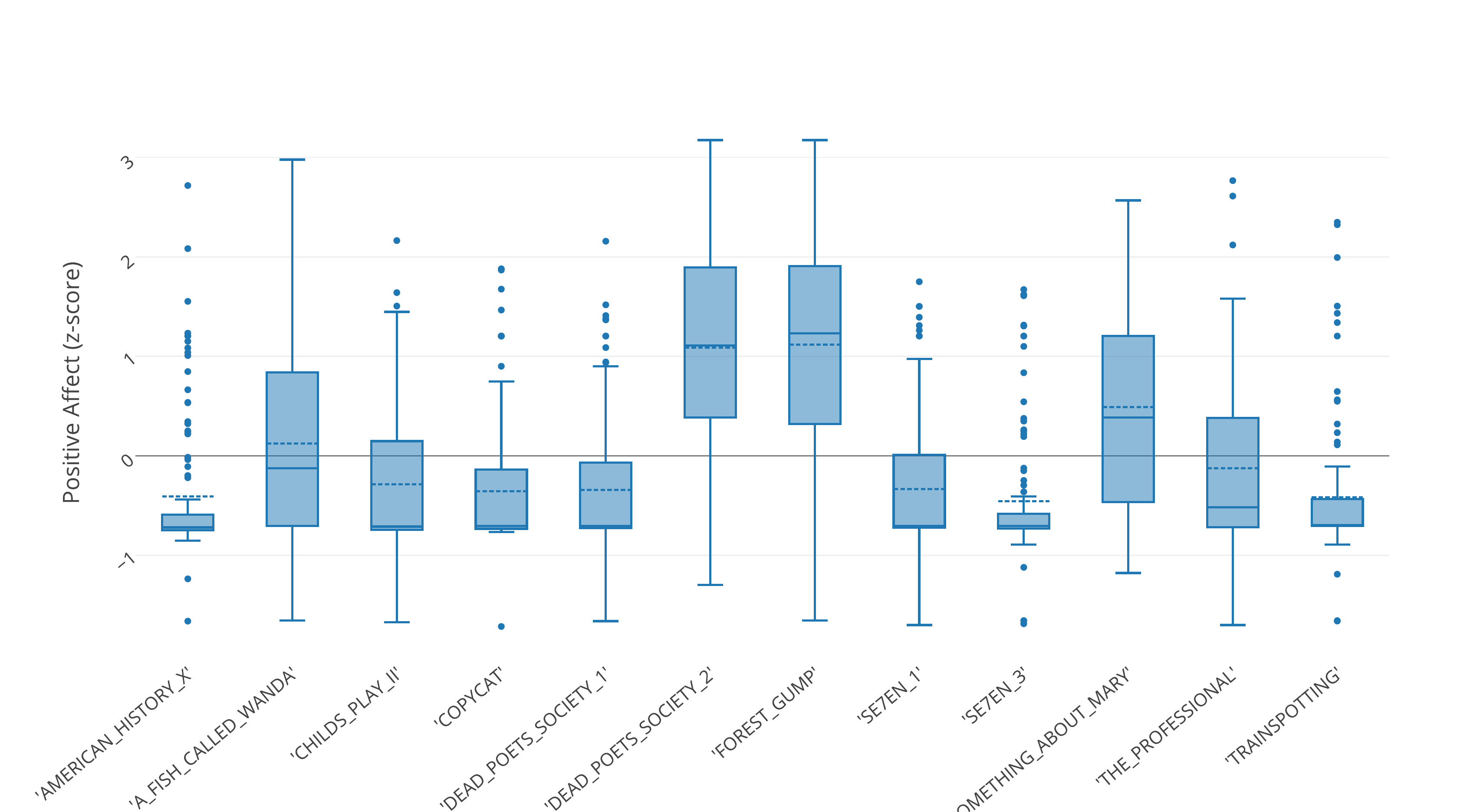}
	\caption{Distribution of positive affect in the dataset. Mean is represented by the dotted line.}
	\label{fig:positive_video_bar}
\end{figure}

\begin{figure}[t]
	\centering
	\includegraphics[width=1\columnwidth]{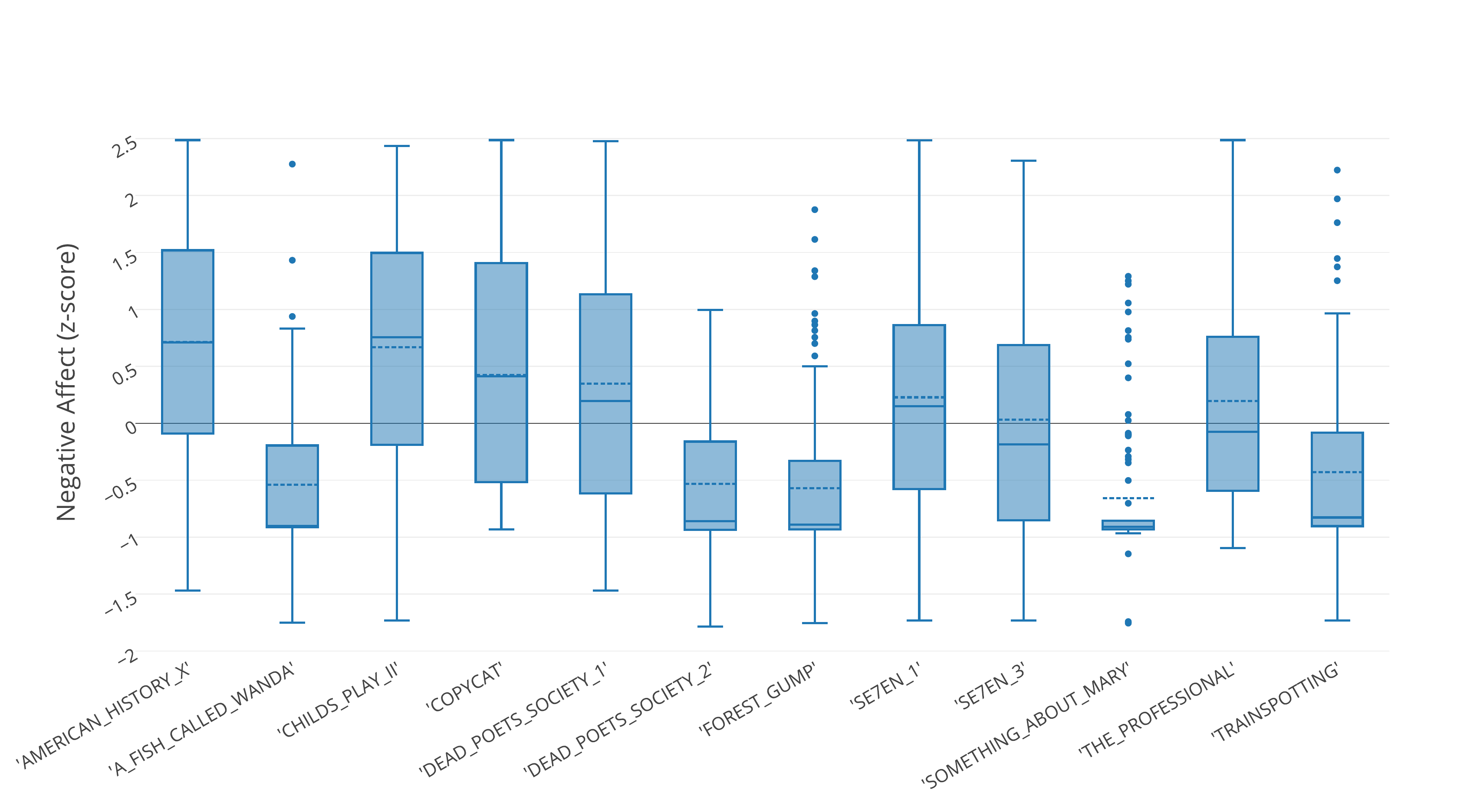}
	\caption{Distribution of negative affect in the dataset. Mean is represented by the dotted line.}
	\label{fig:negative_video_bar}
\end{figure}

was measured using Differential Emotions Scale \cite{mchugo1982structure}. This includes 16 sets of emotion-related adjectives. We refer the reader to \cite{mchugo1982structure} for the list of all sets. Each such set is linked to one of the 5 Likert scales and each participant rated the intensity of felt emotion. The emotions joy, warmth, love, calm, and so on were clubbed as positive affect, and anger, fear, anxiety, sadness, etc. were grouped as negative affect and their aggregate scores were calculated. Descriptive statistics on ratings are shown in Figures \ref{fig:positive_video_bar} and \ref{fig:negative_video_bar}.


\subsubsection{Enjoyment} 
\begin{figure}[t!]
	\centering
	\includegraphics[width=1\columnwidth]{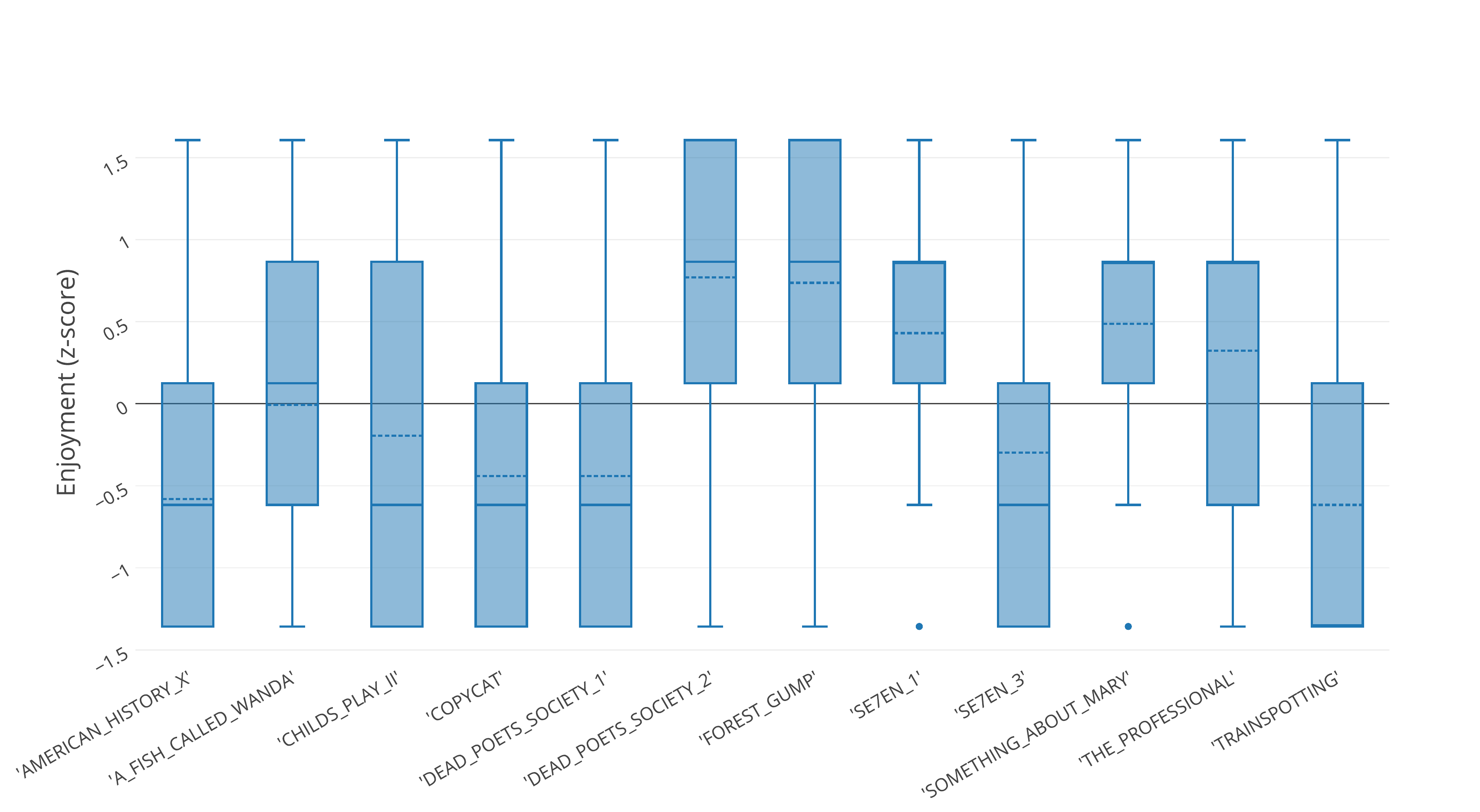}
	\caption{Distribution of enjoyment in the dataset. Mean is represented by the dotted line.}
	\label{fig:enjoyment_bar}
\end{figure}

This was measured using one of the 5-point Guttman-type scales, and each participant indicated how much he/she enjoyed a video sequence.  A value of 1 represents ``no'' enjoyment and a value of 5 denotes ``high'' enjoyment. Descriptive statistics on ratings are shown in Figure \ref{fig:enjoyment_bar}.

\subsubsection{Culture} It was measured using the VSM-2013 questionnaire \cite{hofstede2013values} according to the following aspects: individualism (IDV), power distance (PDI), uncertainty avoidance (UAI), pragmatism (PRG), masculinity (MAS), and indulgence (IVR). We refer the reader to \cite{hofstede2013vsm} for the list of all questions and how they relate to different cultural traits. Most of these questions are on a scale of 1-5.

\subsubsection{Personality} This was measured using the BFI-10 \cite{gosling2003very} questionnaire, according to the FFM \cite{goldberg1990alternative}, measuring conscientiousness (Con), openness (Ope), Extroversion (Ext), Neuroticism (Neu), and Agreeableness (Agr). The questions of BFI-10 along with the corresponding traits is shown in Table \ref{tab:bfi10}. Most of these questions are on a scale of 1-5.

\begin{table}[t!]
	\centering
	\caption{Sample Descriptives.}
	\renewcommand{\arraystretch}{1.2}
	\setlength\tabcolsep{3pt}
	\centering
	\begin{tabulary}{1\columnwidth}{p{2.7cm}rrrrrrr}
		\hline
		\noalign{\smallskip}
		Human Factors & Min & Max & $\bar{x} (NTU)$ & $\bar{x} (BUL)$ & $\bar{x}$ & $\sigma$\\
		\noalign{\smallskip}
		\hline
		\noalign{\smallskip}
		Openess & 4 & 10 &6.60 &6.91 & 6.75 & 1.424 \\
		Conscientiousness & 2 & 10 &6.40 &6.70 & 6.55 & 1.523 \\
		Extroversion & 2 & 9 & 5.61 & 5.46 & 5.54 & 1.689 \\
		Neuroticism & 2 & 10 &5.56 &5.68 & 5.62 & 1.716 \\
		Agreeableness & 3 & 10 &7.33 &7.31 & 7.22 & 1.533 \\
		Individualism & -140 & 140 &25.79 &11.67 & 18.73 & 50.619 \\
		Power Distance & -155 & 140 &-35.61 &-36.32 & -35.96 & 53.219 \\
		Masculinity & -140 & 105 &3.68 &-6.14 & -1.23 & 53.483 \\
		Pragmatism & -130 & 155 &16.14 &17.54 & 16.84 & 58.090 \\
		Uncertainty Avoidance & -120 & 130 &52.54 &36.67 & 44.61 & 47.182 \\
		Indulgence & -220 & 185 &-22.63 &-11.32 & -16.97 & 65.522 \\ \hline
	\end{tabulary}
	\label{tab:descstat}
\end{table}


\section{Methodology}

In this section, we introduce different statistical mothods used, the features extracted to build the predictive models and the evaluations. 

\subsection{Statistical Analysis}
The analysis has been conducted in PASW 20.0. Linear mixed-effects modeling has been adopted for repeated measures, with model parameters determined with the restricted maximum-likelihood method. 

We build three computational models (namely baseline, extended and optimistic) to investigate the influence of system factors -- framerate, bitrate and resolution, human factors -- five personality factors and six culture factors, on the experience of affect and enjoyment. Each of them along with corresponding findings will be described in this section. Afterwards a comparison between the three models will be presented to address the four questions we pose in this paper.

\subsubsection{Baseline Model}
This model considers only system factors. For the CP-QAE-I video dataset, there are 12 variations of the system factors -- framerate , bitrate, and resolution/framesize. Factors such as format of the files and network protocol were held constant. Due to the expected interactions between these conditions (e.g., an attempt to minimise Bit-Rate while maximising Frame-Rate and framesize would likely create artefacts), they have been modelled as factorial interactions. In addition, the movie clip itself is included as a parameter to reflect differences in cinematographic schemes used to create the movies, along with the nature of the content. This was modelled as a main effect.

\subsubsection{Extended Model}
The extended model adds additional fixed parameters to the baseline model. These were cultural traits: individualism, power distance, masculinity, pragmatism, uncertainty avoidance, and indulgence. Additionally, personality traits were also added: extroversion, agreeableness, conscientiousness, neuroticism, and openness. These were incorporated into the model as covariates with direct effects.


\subsubsection{Optimistic Model}
While a model aims at predicting a dependent variable as precisely as possible, not all of the residual variance can be solely attributed to human factors. A non-trivial proportion of the residual variance can also, to name but a few, be attributed to random error, measurement error, and the limitations of the modelling technique (in this case, generalised linear regression). As such, an optimistic model can be used to estimate the part of the residual variance that might possibly be attributed to human factors in general and, to a small extent, because of the factors such as context and limitations in experimental control. This is achieved by modelling every participant as a random effect, i.e.,  measurements have been repeated to obtain a different intercept during the regression for every participant.

\subsection{Predictive Modeling}
While statistical analysis provides an understanding of the relationships of different dependent and independent variables in the data, predictive models help to forecast users' responses on new data samples. 

We propose a prediction framework which takes input features in video content, system characteristics, personality and cultural traits, predicting experience of both positive and negative affect, and enjoyment on the video clips using $L_{1}$ regularised $L_2$-loss sparse SVM with a linear kernel. We chose a linear kernerl to avoid the problem of overfitting (as seen in literature \cite{borth2013sentibank, donahue2013decaf}). We use the libsvm framework \cite{chang2011libsvm} to carry out these experiments.

\subsubsection{Features}

We use 4 categories of features to represent content-based, system-based, affective and human factors, in an attempt to describe various features which might influence the experience of affect and enjoyment. They are described as follows.

\paragraph*{Content-factors} To represent the different concepts in the video, we used HybridCNN \cite{zhou2014learning}, Adjective-Noun Pairs \cite{borth2013sentibank}, Color histograms, Bag-of-Visual-Words \cite{yang2007evaluating}, Aesthetic features \cite{bhattacharya2013towards, machajdik2010affective} and LBP features \cite{ojala2002multiresolution}.  Representation for every video is gotten by mean-pooling on features obtained from all frames. 

\textbf{Color Histogram}: Users' perception is greatly influenced by the color hues in the videos. Therefore, color histograms are chosen to represent users' color inclination.

\textbf{Aesthetic Features}: We used 2 sets of features to represent the aesthetic charactersitics in videos: a.)  art and psychology based features \cite{machajdik2010affective} to describe photographic styles (rule-of-thirds, vanishing points, etc.). b.) psycho-visual characteristics at cell, frame  and shot levels proposed by \cite{bhattacharya2013towards}. 

\textbf{LBP}: LBP was used to encode users' percetion of texture in the videos. As LBP represents facial information well and many of the videos have people, we use LBP features.

\textbf{Bag-of-Visual-Words} \cite{yang2007evaluating}: A 1500 dimension feature based on vector quantization of keypoint descriptors is generated for every frame. A bag of visual words representation is used by mapping the keypoints to visual words.  

\textbf{HybridPlacesCNN} : CNN features from fc7-ReLu layer from ImageNet \cite{krizhevsky2012imagenet} and Places dataset \cite{zhou2014learning} are used to represent objects and places in videos. 

\textbf{Adjective-Noun Pairs (ANP)}: Emotion based attributes are detected (2089 ANPs)  \cite{borth2013sentibank} using the sentibank classifiers. 8 emotion categories \cite{plutchik1980emotion} are used to define the adjectives, and objects and scenes are used to define the nouns. 

\begin{table*}[t!]
	\centering
	\caption{List of Features}
	\label{tab:featall}
		\renewcommand{\arraystretch}{1.2}
		\setlength\tabcolsep{3pt}
	\begin{tabular}{c|c|L{10cm}}
		\hline
		\textbf{Category}                  & \textbf{Feature}                 & \textbf{Description}                                                                                                                                                      \\ \hline
		\multirow{6}{*}{Visual Content}    & Color Histogram                  & based on RGB values of each frame                                                                                                                                         \\
		& Visual Aesthetics                & art and psychology based features to describe photographic styles (rule-of-thirds, vanishing points, etc.); psycho-visual characteristics at cell, frame  and shot levels \\
		& LBP                              & users' percetion of texture in the videos                                                                                                                                 \\
		& BoVW                             & 1500-dimension vector based on quantization of keypoint descriptors                                                                                                       \\
		& HybridPlaces CNN                 & features from the ReLu layer following fc7 layer of a CNN trained on 1.3 million images from  ImageNet  and 2.5 million images on the Places dataset                      \\
		& Sentibank                        & detection of 2089 Adjective Noun Pairs based on emotion related concepts                                                                                                  \\ \hline
		\multirow{3}{*}{Audio Content}     & Musical chroma                   & \multirow{2}{*}{Opensmile was used to extract the affective features from audio signal}                                                                                   \\
		& Prosody                          &                                                                                                                                                                           \\
		& Low-level descriptors            & intensity, MFCC, loudness,  pitch, pitch envelope, probability of voicing, zero-crossing rate and line spectral frequencies                                               \\ \hline
		\multirow{2}{*}{System parameters} & Bit Rate, Frame Rate, Resolution &                                                                                                                                                                           \\
		& Perceptual Characteristics       &  Quality of the distorted
		image is expressed as a simple distance metric between the
		model statistics and those of the distorted image                                                                                                                                                                         \\ \hline
		\multirow{3}{*}{Human factors}     & Personality                      & openness, conscientiousness, extraversion, agreeableness, neuroticism                                                                                                     \\
		& Cultural traits                  & power distance, masculinity, individualism, uncertainity avoidance, indulgence, pragmatism                                                                                \\
		& Demography                       & gender, age and nationality                                                                                                                                               \\ \hline
	\end{tabular}
\end{table*}

\paragraph*{Audio Affect-factors} Affective characteristics associated with the audio content in videos are extracted (OpenSmile \cite{eyben2010munich}) as musical chroma features \cite{muller2011chroma}, prosodic features \cite{carey1996robust} and low-level descriptors such as MFCC features, intensity, loudness,  pitch,  pitch envelope, probability of voicing, line spectral frequencies, and zero-crossing rate. The visual affective content in the videos was expected to be represented by the aesthetic and ANP features described above.

\paragraph*{System-factors} Bitrate, Framerate, resolution and perceptual characteristics \cite{mittal2013making} are used to represent quality characteristics of videos. Perceptual characteristics describe the no-reference quality metric \cite{mittal2012no}. Perceptual characteristics were represented by temporal distortions in the video, spatial domain natural scene statitstics,  statistical DCT features motion coherence feature describing the coherence in strength and direction of local motion due to temporal distortions, and reflecting the perceptural difference between pristine and distorted videos. 

\paragraph*{Human-factors} The five personality factors, six cultural traits, gender, age and nationality of the users are used to represent human factors. 

\section{Results and Discussion}
Results target at answering the four research questions raised at the outset of the paper. Sections \ref{subsec:models} and \ref{subsec:comparison} deal with RQ1 and RQ2, Section \ref{subsec:correl} deals with RQ3, and  Section \ref{subsec:pred} answers RQ4.

\subsection{Statistical Analysis}
\label{subsec:models}

\begin{table*}[t!]
	\centering
	\caption{Baseline fixed-effect multilevel linear regression model}
	\renewcommand{\arraystretch}{1.3}
	\setlength\tabcolsep{3pt}
	\begin{tabulary}{1\columnwidth}{lrrrrrrrrrrrrrr}
		\hline \noalign{\smallskip}
		& \multicolumn{2}{p{1.2cm}}{}		& \multicolumn{3}{c}{Positive Affect}		&& \multicolumn{3}{c}{Negative Affect} 			&& \multicolumn{3}{c}{Enjoyment}	\\ 
		\cline{4-6} \cline{8-10} \cline{12-14}
		Parameter                                				& $df_{num}$ 	&& $df_{den}$ 	& $F$      	& $p$ 			&& $df_{den}$ 	& $F$      	& $p$ 			&& $df_{den}$ 	& $F$      	& $p$ 		\\ 
		\noalign{\smallskip}
		\hline
		\noalign{\smallskip}
		Movie Clip      & 11 &  & 156.009 & 25.315 & 0.00 &  & 144.643 & 33.932 & 0.00 &  & 177.09   & 40.14 & 0.00     \\
		Frame Rate (FR) & 2  &  & 803.739 & 0.32   & 0.73 &  & 710.192 & 0.056   & 0.95 &  & 1131.23  & 5.173 & 0.006 \\
		Frame Size (FS) & 1  &  & 809.889 & 0.006   & 0.94 &  & 729.398 & 3.298  & 0.07 &  & 1146.39  & 2.846 & 0.092 \\
		Bit-Rate (BR)   & 1  &  & 816.675 & 1.724  & 0.19 &  & 714.909 & 0.30   & 0.58 &  & 1139.69  & 0.474 & 0.491 \\
		\hline
	\end{tabulary}
	\small
	\\
	\vspace{1mm}
	\centering{\textit{Interactions of System Factors namely FR $\times$ FS, FS $\times$ BR, FR $\times$ BR, FR $\times$ FS $\times$ BR were found to be insignificant predictors and hence not included in the above table.}}
	\label{tab:baseline}
\end{table*}

\begin{table*}[t!]
	\centering
	\caption{Extended fixed-effect multilevel linear regression model}
	\renewcommand{\arraystretch}{1.3}
	\setlength\tabcolsep{3pt}
	\begin{tabulary}{1\columnwidth}{lrrrrrrrrrrrrrr}
		\hline \noalign{\smallskip}
		& \multicolumn{2}{p{1.2cm}}{}		& \multicolumn{3}{c}{Positive Affect}		&& \multicolumn{3}{c}{Negative Affect} 			&& \multicolumn{3}{c}{Enjoyment}	\\ 
		\cline{4-6} \cline{8-10} \cline{12-14}
		Parameter                                				& $df_{num}$ 	&& $df_{den}$ 	& $F$      	& $p$ 			&& $df_{den}$ 	& $F$      	& $p$ 			&& $df_{den}$ 	& $F$      	& $p$ 		\\ 
		\noalign{\smallskip}
		\hline
		\noalign{\smallskip}
		Movie Clip            & 11 &  & 193.163  & 35.925 & 0.00 &  & 206.260  & 39.739 & 0.00 &  & 171.956  & 39.733 & 0     \\
		Frame Rate (FR)       & 2  &  & 1071.695 & 0.18   & 0.84 &  & 1045.660 & 0.48   & 0.62 &  & 1136.577 & 4.695  & 0.009 \\
		Frame Size (FS)       & 1  &  & 1074.152 & 0.54   & 0.46 &  & 1061.874 & 2.10  & 0.15 &  & 1151.402 & 3.336  & 0.068 \\
		Bit-Rate (BR)         & 1  &  & 1083.535 & 2.334  & 0.13 &  & 1044.851 & 0.06   & 0.807 &  & 1145.171 & 0.257  & 0.612 \\
		Extraversion          & 1  &  & 1074.324 & 4.559  & 0.033 &  & 1059.767 & 0.08   & 0.78 &  & 1150.401 & 0.024  & 0.877 \\
		Agreeableness         & 1  &  & 1072.223 & 1.876  & 0.17 &  & 1059.481 & 24.314 & 0.00 &  & 1152.475 & 2.001  & 0.157 \\
		Conscientiousness     & 1  &  & 1077.950 & 9.474  & 0.002 &  & 1041.655 & 3.964  & 0.047 &  & 1141.249 & 5.271  & 0.022 \\
		Neuroticism           & 1  &  & 1084.026 & 0.02   & 0.888 &  & 1050.845 & 25.227 & 0.00 &  & 1146.479 & 0.05   & 0.823 \\
		Openness              & 1  &  & 1074.213 & 2.670  & 0.103 &  & 1058.628 & 2.110  & 0.147 &  & 1145.365 & 4.344  & 0.037 \\
		Power Distance        & 1  &  & 1073.888 & 4.676  & 0.031 &  & 1055.500 & 0.00   & 0.985 &  & 1152.465 & 9.138  & 0.003 \\
		Individualism         & 1  &  & 1070.708 & 2.148  & 0.143 &  & 1052.462 & 2.486  & 0.115 &  & 1150.026 & 0.674  & 0.412 \\
		Masculinity           & 1  &  & 1074.304 & 4.874  & 0.027 &  & 1043.258 & 1.061  & 0.303 &  & 1141.312 & 3.312  & 0.069 \\
		Uncertainty Avoidance & 1  &  & 1077.284 & 0.534   & 0.465 &  & 1044.360 & 0.306   & 0.580 &  & 1144.106 & 5.751  & 0.017 \\
		Pragmatism            & 1  &  & 1069.661 & 0.886   & 0.347 &  & 1064.578 & 0.175   & 0.676 &  & 1160.7   & 0.604  & 0.437 \\
		Indulgence            & 1  &  & 1070.162 & 5.863  & 0.016 &  & 1051.545 & 4.863  & 0.028 &  & 1149.178 & 2.206  & 0.138 \\ \hline
	\end{tabulary}
	\small
	\\
	\vspace{1mm}
	\centering{\textit{Interactions of System Factors namely FR $\times$ FS, FS $\times$ BR, FR $\times$ BR, FR $\times$ FS $\times$ BR were found to be insignificant predictors and hence not included in the above table.}}
	\label{tab:extended}
\end{table*}

\subsubsection{Baseline Model}
Table \ref{tab:baseline} shows the results. The movie clip itself had the largest impact on experience of affect and enjoyment. However, an interesting observation is that only Frame Rate had a statistically significant effect on enjoyment. This shows that system factors alone do not make a huge impact on how the content is perceived. That is, given two videos of different natures at different bitrate, resolution/framesize and framerate, the nature of the content alone is more likely to influence how it is perceived than the system settings at which it is delivered. Our findings can be corroborated by similar observations in QoE \cite{ghinea1998qos,yeung2004independent}.

\begin{table*}[t!]
	\centering
	\caption{Optimistic mixed-effect multilevel linear regression model}
	\renewcommand{\arraystretch}{1.3}
	\setlength\tabcolsep{3pt}
	\begin{tabulary}{1\columnwidth}{lrrrrrrrrrrrrrr}
		\hline \noalign{\smallskip}
		& \multicolumn{2}{p{1.2cm}}{}		& \multicolumn{3}{c}{Positive Affect}		&& \multicolumn{3}{c}{Negative Affect} 			&& \multicolumn{3}{c}{Enjoyment}	\\ 
		\cline{4-6} \cline{8-10} \cline{12-14}
		Parameter                                				& $df_{num}$ 	&& $df_{den}$ 	& $F$      	& $p$ 			&& $df_{den}$ 	& $F$      	& $p$ 			&& $df_{den}$ 	& $F$      	& $p$ 		\\ 
		\noalign{\smallskip}
		\hline
		\noalign{\smallskip}
		Movie Clip      & 11 &  & 178.713 & 42.312 & 0.00  &  & 152.624 & 55.782 & 0.00 &  & 179.877  & 46.99 & 0.00  \\
		Frame Rate (FR) & 2  &  & 701.036 & 1.788  & 0.168  &  & 945.140 & 1.392  & 0.249 &  & 1116.89  & 8.025 & 0.00  \\
		Frame Size (FS) & 1  &  & 695.825 & 0.002   & 0.965  &  & 969.366 & 5.764  & 0.017 &  & 1120.818 & 3.13  & 0.077 \\
		Bit-Rate (BR)   & 1  &  & 715.664 & 1.159  & 0.282  &  & 972.050 & 1.457  & 0.228 &  & 1121.96  & 0.054 & 0.816 \\
		\hline
	\end{tabulary}
	\small
	\\
	\vspace{1mm}
	\centering{\textit{Interactions of System Factors namely FR $\times$ FS, FS $\times$ BR, FR $\times$ BR, FR $\times$ FS $\times$ BR were found to be insignificant predictors and hence not included in the above table.}}
	\label{tab:optimistic}
\end{table*}

\subsubsection{Extended Model}
Table \ref{tab:extended} shows the results of the extended model. Many personality and cultural traits seem to be significant predictors of experience of affect and enjoyment. Among personality traits, extraversion and conscientiousness are significant predictors for positive affect, and agreeableness, neuroticism and conscientiousness are significant for negative affect. Conscientiousness and openness are significant predictors for enjoyment \cite{ izard1993stability, pessoa2008relationship}. Among cultural traits, masculinity and indulgence are significant predictor for positive affect, indulgence alone for negative affect and uncertainity avoidance for enjoyment. None of the system factors (except Frame-Rate for enjoyment) and their interactions are significant predictors.

This suggests that 	multimedia system characteristics have little to no influence on the intensity of the affect that viewers experience. Additionally, there appears to be a different set of predictors for affect compared to overall enjoyment. $F$-statistic is generally quite small for most of the predictors. However, the predictors of agreeableness and neuroticism, for negative affect, are notably much larger. This suggests that a considerable amount of the variance in negative affect can be explained by these parameters.

\subsubsection{Optimistic Model}
Table \ref{tab:optimistic} shows the results of the optimisitic models. The model is quite similar to the baseline model with the exception of larger $F$-statistics, indicating that larger proportion of variance is explained because of considering random effects. Additionally, interaction between framesize and experience of affect is now significant. 

\subsubsection{Model Comparison}
\label{subsec:comparison}
Paired t-tests on Mean Squared Residuals (MSR) are used to compare the models as shown in Table  \ref{tab:modelimprove}. Proportional reduction in overall MSE is shown (see \cite{bosker2012multilevel}). The results show that human factors, namely personality and culture, play a crucial role in modeling the experience of affect and enjoyment, indicating that content production and delivery mechanisms should not just take into account multimedia system factors but also human factors to achieve maximal user satisfaction.

\paragraph*{Models for Positive Emotion}
From the baseline to optimistic model, the MSR reduced from 0.6304 ($\sigma = 1.050$) to 0.4051 ($\sigma = 0.886$ ; $p < 0.005$), representing a predicted variance of 55.3\%. A part of this is contributed by culture and personality. 5.6\% of variance attributable to human factors is predicted by the extended model, reducing the baseline MSR to 0.6177 ($\sigma = 1.005$ ; $p = 0.021$).

\paragraph*{Models for Negative Emotion}
From the baseline to optimistic model, the MSR reduced from 0.6514 ($\sigma = 0.889$) to 0.3615 ($\sigma = 0.536$ ; $p < 0.00$), representing a predicted variance of 58.1\%. 13.6\% of variance attributable to human factors, reducing the baseline MSR to 0.6118 ($\sigma = 0.8278$ ; $p < 0.00$).

\paragraph*{Models for Enjoyment}
From the baseline to optimistic model, the MSR reduced from 1.3684 ($\sigma = 1.63$) to 0.9481 ($\sigma = 1.22$ ; $p < 0.00$), which makes up 23.0\% of the overall variance predicted. 9.3\% of variance due to human factors is predicted by the extended model, which decreases the baseline MSR to 1.3290 ($\sigma = 1.58$ ; $p < 0.001$).

\begin{table*}[t!]
	\centering
	\caption{Paired t-test showing the comparison of models for all three responses (w.r.t MSR)}
	\renewcommand{\arraystretch}{1.3}
	\setlength\tabcolsep{3pt}
	\begin{tabulary}{1\columnwidth}{lrrrrrrrrrrrrrrrr}
		\hline \noalign{\smallskip}
		& \multicolumn{1}{c}{}		& \multicolumn{4}{c}{Positive Affect}		&& \multicolumn{4}{c}{Negative Affect} 			&& \multicolumn{4}{c}{Enjoyment}	\\ 
		\cline{3-6} \cline{8-11} \cline{13-16}
		Models  && $\Delta \bar{x}$ & $\sigma$ & $t$ & $p$ 	&& $\Delta \bar{x}$ & $\sigma$ & $t$ & $p$ && $\Delta \bar{x}$ & $\sigma$ & $t$ & $p$ \\ 
		\noalign{\smallskip}
		\hline
		\noalign{\smallskip}
		$Baseline \rightarrow Extended$    && 0.013 & 0.193 & 2.311 & 0.021 && 0.039  & 0.277 & 5.008 & 0.00  && 0.039  & 0.430  & 3.219 & 0.001 \\
		$Baseline \rightarrow Optimistic$  && 0.2253 & 0.924 & 8.552 & 0.00  && 0.2898  & 0.726 & 14.014& 0.00  && 0.4199  & 1.129 & 13.069& 0.00 \\
		
		\hline
	\end{tabulary}
	\label{tab:modelimprove}
\end{table*}

\begin{table*}[tp]
	\centering
	\caption{Significant correlations between enjoyment and experience of affect ($p<0.05$)}
	\label{tab:corr_enjaff}
	\renewcommand{\arraystretch}{1.3}
	\setlength\tabcolsep{3pt}
	\begin{tabulary}{1\columnwidth}{l|rrrrrrrrrrrrrrr}
		\hline \noalign{\smallskip}
		Clip & Interest & Joy  & Sad   & Fearful & Disgust & Surprise & Warm & Loving & Guilty & Moved & Satisfied & Calm & Ashamed & +ve Affect \\ \hline
		C-I    & 0.579     & 0.699 & -    & -       & -0.332   & 0.430    & 0.626 & 0.419   & -      & -     & 0.456      & 0.451 & -       & 0.610              \\
		C-II    & 0.505     & 0.304 & -    & -       & -0.332   & 0.380    & 0.350 & 0.383   & 0.258   & -     & 0.526      & 0.404 & 0.291    & 0.483              \\
		C-III    & 0.596     & 0.485 & -    & -       & -0.314   & 0.250    & 0.298 & 0.247   & -      & -     & 0.432      & 0.286 & -       & 0.426              \\
		C-IV    & 0.444     & 0.424 & -    & -       & -       & -       & 0.287 & 0.243   & -      & -     & 0.255      & -    & -       & 0.304              \\
		C-V    & 0.469     & 0.263 & -     & -       & -       & -       & 0.368 & 0.271   & 0.244   & 0.219  & 0.288      & 0.232 & 0.250    & 0.298              \\
		C-VI    & 0.514     & 0.385 & -    & -0.239   & -       & -       & 0.353 & 0.325   & -      & -     & 0.493      & -    & -0.215   & 0.456              \\
		C-VII    & 0.549     & 0.643 & 0.248 & -       & -       & 0.293    & 0.479 & 0.513   & -      & 0.508  & 0.510      & 0.407 & -       & 0.560              \\
		C-VIII    & 0.550     & 0.408 & -    & -       & -       & -       & -    & -      & -      & 0.346  & 0.445      & 0.319 & -       & 0.421              \\
		C-IX    & 0.340     & 0.394 & -    & -       & -       & -       & 0.323 & 0.251   & -      & -     & 0.219      & 0.244 & -       & 0.292              \\
		C-X   & 0.512     & 0.658 & -    & -       & -       & 0.258    & 0.301 & -      & -      & -     & 0.267      & 0.354 & 0.290    & 0.419              \\
		C-XI   & 0.541     & 0.266 & -    & -       & -       & 0.232    & -    & -      & -      & 0.277  & 0.358      & 0.325 & -       & 0.347              \\
		C-XII   & 0.590     & 0.688 & -    & -       & -0.401   & 0.317    & 0.434 & 0.419   & 0.244   & 0.295  & 0.488      & 0.353 & -       & 0.542            \\
		\hline
	\end{tabulary}
	\small
	\\
	\vspace{.2cm}
	\raggedright{\centering\textit{{Movie Clips:- C-I: A\_FISH\_CALLED\_WANDA; C-II: AMERICAN\_HISTORY\_X; C-III: CHILDS\_PLAY\_II; C-IV: COPYCAT; C-V: DEAD\_POETS\_SOCIETY\_1; C-VI: DEAD\_POETS\_SOCIETY\_2; C-VII: FOREST\_GUMP; C-VIII: SE7EN\_1; C-IX: SE7EN\_3; C-X: SOMETHING\_ABOUT\_MARY; C-XI: THE\_PROFESSIONAL; C-XII: TRAINSPOTTING. No significant correlations were observed between Enjoyment and Anxious, Angry, Disdain \& -ve Affect and thus are not shown in the table. Entry $E_{i,j}$ represents the correlation between enjoyment and affect for movie clip $i$ and emotion category $j$ }}}
	
\end{table*}

\subsubsection{Correlation between Affect and Enjoyment}
\label{subsec:correl}
As introduced at the beginning of the article, there is a very close and significant relationship between what users enjoy and the emotion it evokes (results from correlation analysis are shown in Table \ref{tab:corr_enjaff}). In all clips, enjoyment is significantly correlated with interest, joy, satisfaction and the latent factor, positive emotion. This means that for a user to enjoy a video the content has to definitely draw his/her interest, but must also have moments of happiness and deliver something which satisfies the viewer \cite{soto2011influence}. 

There are also very few instances of negative emotions (sad, fearful, guilty, and ashamed) giving enjoyment to users. For instance, enjoyment was seen to have a significant positive correlation with emotions Ashamed and Guilty for both the clips AMERICAN\_HISTORY\_X (in which the protagonist is seen brutally torturing someone) and DEAD\_POETS\_SOCIETY (in which one of the main characters commits suicide by shooting himself) and with emotion Ashamed for SOMETHING\_ABOUT\_MARY (in which there is an obscenity involved and yet sounds joyful/funny) and Guilty for TRAINSPOTTING (in which a person is seen to get inside a dirty toilet bowl, and yet the music is of a totally different contrast). The predominant emotion in these clips are not widely enjoyable. However, these might be associated with how certain users (possibly with high scores on neuroticism) perceive certain contents \cite{diener2003personality, larsen1991personality}

Apart from that, even the nature of the content itself can arouse contradictory emotions. For example, enjoyment is observed to be positively correlated with sadness in the movie clip FOREST\_GUMP which has a defining excerpt in which the leading character encounters his son for the very first time. This is a scene with bitter-sweet connotations for viewers due to the fact that the protagonist, quite belatedly in his life, is faced with the news that not only has he fathered a son, but also that the son is doing well in school and is a fine student. So, such occurrences are due to the interaction of human factors and nature of the content both.

\begin{figure}[t]
	\centering
	\includegraphics[width=1\columnwidth]{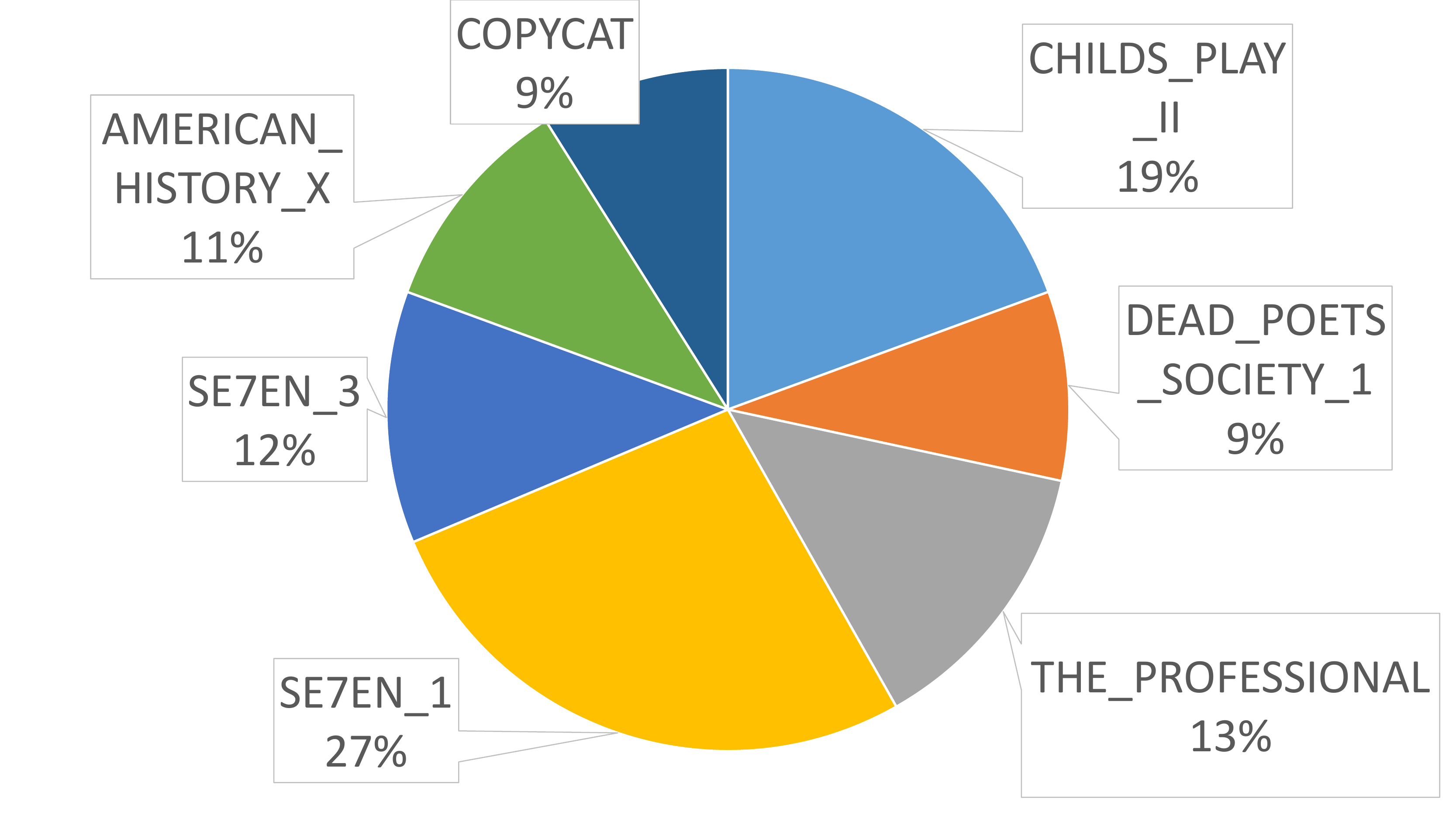}
	\caption{Distribution of ratings on movie clips with high enjoyment and high -ve affect.}
	\label{fig:pie}
\end{figure}

\begin{figure*}[t!]
	\centering
	\includegraphics[width=0.8\textwidth]{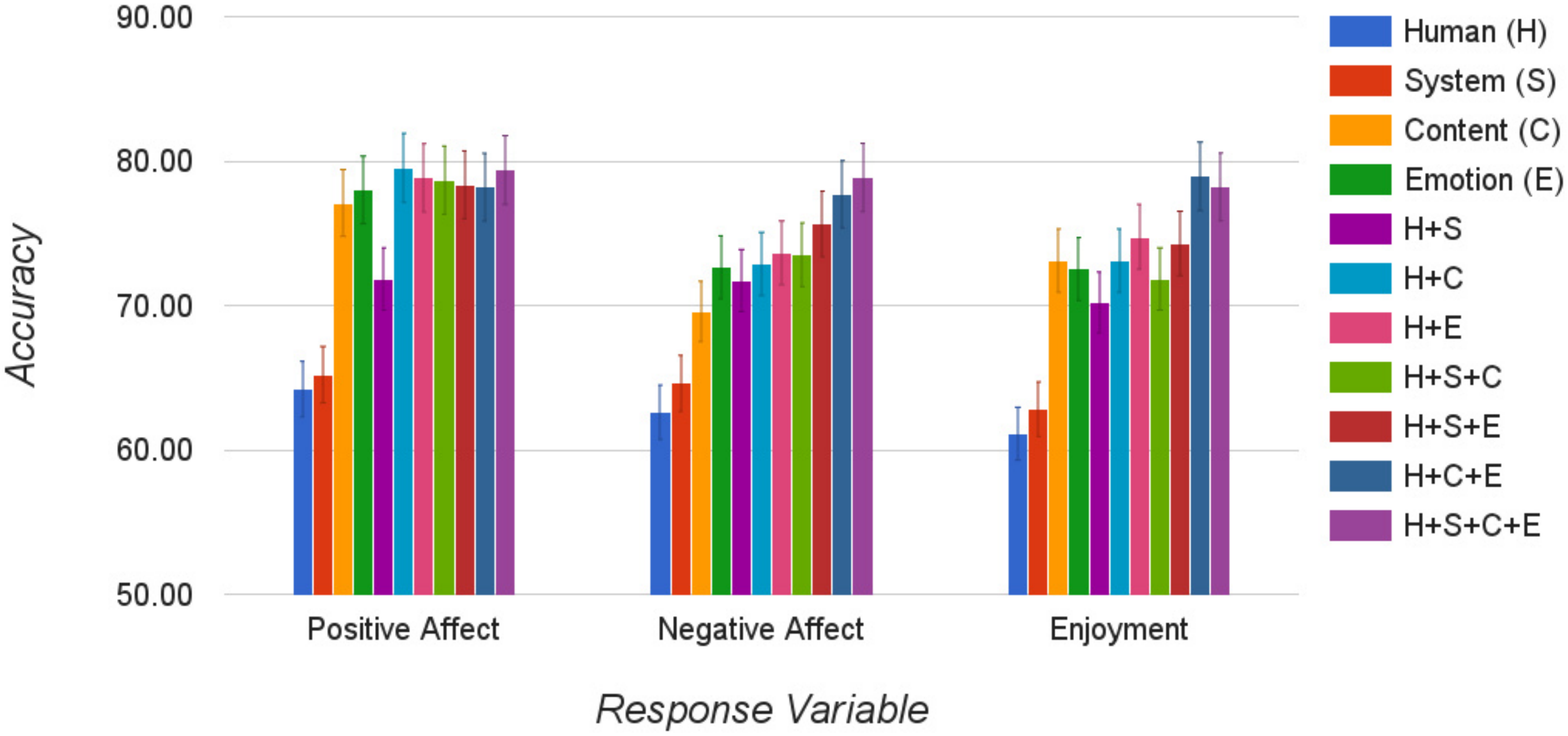}
	\caption{Predictive models' accuracy when employing different features.}
	\label{tab:pred}
\end{figure*}

It is interesting to note that while most of the users might associate enjoyment with positivity, there are certain users who need to experience negative emotions to connect to the content's message. This insight gives content creators a better understanding of how to influence users with different personality and cultural traits to establish an emotional connection with them, which is very important to drive behavioral action (especially in scenarios involving ad campaign design etc.).

To investigate this further, we selected the records with high enjoyment (i.e., 4 and 5) and with high negative affect (above mean) and low positive affect (below mean). The filtered subset (about 5.4\% of the total ratings) was investigated to understand the nature of content which likely influences such rating behavior. Figure \ref{fig:pie} shows distribution of clips in the filtered subset.

\subsection{Predictive Models}
\label{subsec:pred}

Experiments are run on leave-one-video-out setting by binarising perceptual quality and enjoyment scores. Figure \ref{tab:pred} depicts the accuracy with which the trained models predict user affect and enjoyment. Human, system, content and emotion factors (the latter based on audio affect) were employed, and features extracted to represent the same.

As far as positive and negative affect are concerned, models which gave the best performance were those trained on emotion and content factors (in this order). This is intuitive, as the nature of content plays the most important role in influencing viewers' experience of affect. It was also seen in our statistical analysis (in the previous section).

It was then explored if adding human factors to those pertaining to system, content and emotion  would improve the predictive modeling performance. 

In terms of positive affect, the combination of human and content, followed by that of human and emotion factors yielded the best results. It must be noted that the features representing content factors also include ANPs, which are especially designed for visual sentiment prediction \cite{borth2013sentibank}. This explains their almost equivalent performance when compared to emotion factors. Similar performance was seen when three factors (namely content, human and emotion) are combined to train the model. Similar observations are made for negative affect, however the performance was lower than that of predicting positive affect. This is similar to the observations made in other works \cite{gunes2011emotion, jou2014predicting, sanchez2013inferring,  siddiquie2015exploiting, valstar2013avec}, possibly due to the intrinsic challenge of modeling the nature of negative emotions.

As far as enjoyment is concerned, models trained on human, emotion, and content factors performed better than others. Our statistical analysis showed evidence of this, as content was found to be significantly correlated with enjoyment . The impact of factors associated with emotion has been explored in other studies\cite{bardzell2009understanding,bilandzic2011enjoyment,visch2010emotional}. 

\section{Conclusion}

Experience of affect and enjoyment in multimedia is influenced by an intricate interplay between characteristics of stimuli, individuals, and systems of perception. Returning to the research questions posed at the outset of the paper, we can now state that:
\begin{enumerate}
	\item [RQ1] For positive affect, negative affect and enjoyment, personality and culture respectively represented  5.6\%, 13.6\% and 9.3\% of variance. Notwithstanding the fact that these constitute sizeable proportions, follow up studies need to explore other potential contributing factors, such as sensory impairnments/acuity, user cognitive style, and domain expertise \cite{ghinea2006perceived}.
	\item [RQ2] Traits of extraversion, conscientiousness, masculinity and indulgence are significant predictors for positive affect, and agreeableness, neuroticism, conscientiousness and indulgence were important predictors for negative affect. Conscientiousness, openness and uncertainity avoidance were significant predictors for enjoyment.
	\item [RQ3] The majority of the movie clips which were enjoyed were also rated high on positive affect, with a small exception of clips having high correlation between negative affect and enjoyment. Such behavior is possibly due to the interchange that potentially takes place between human factors (e.g. neuroticism) and media content. 
	\item [RQ4] Predictive models trained with a mixture of human factors and content, emotion and emotion factors yielded the highest achievement for positive affect, negative affect and enjoyment respectively with an accuracy of 79\%, 77\% and 76\% respectively. 
\end{enumerate} 

It is important to know the impact of human factors on user enjoyment, as this allows one to optimise this latter parameter especially in conditions in which other more traditional forms of adaptation (such as layered adaptation) are difficult/impractical to perform. As highlighted above, results obtained in our study showcase the important part that human factors have on two impartant aspects of user QoE, namely affect and enjoyment. Thus, integration of human factors in the optimistic model was shown to have significantly improved modeling performance; however, extended models based on personality and culture did not have impacts of the same magnitude. This means that several other human factors, apart from those taken into consideration in this study such as user behavior in a particular domain of study (e.g. movies rated or images liked) or other psychological constructs like mood etc. \cite{boll2007multitube, madden1996seasonal, o2013characterizing} can be explored. Nonetheless, results show that human factors, namely personality and culture, exert an important influence in  modeling the experience of affect and enjoyment, indicating that content production and delivery mechanisms should not just take into account multimedia system factors but also human factors, in order to achieve maximal user satisfaction. 

\bibliographystyle{plain}
\bibliography{egbib}
\ifCLASSOPTIONcaptionsoff
  \newpage
\fi

\end{document}